\documentclass[11pt]{article}
\usepackage{epsfig}
\usepackage{amsfonts}
\usepackage{amsmath}
\usepackage{bbm,bm}
\usepackage{cite}
\allowdisplaybreaks[1]

\newcommand{\nr}[1]{(\ref{#1})}
\newcommand{\tr}{{\rm Tr\,}}

\newcommand{\la}[1]{\label{#1}}
\newcommand{\be}{\begin{equation}}
\newcommand{\ee}{\end{equation}}
\newcommand{\ba}{\begin{eqnarray}}
\newcommand{\ea}{\end{eqnarray}}
\newcommand{\rmi}[1]{{\mbox{\scriptsize #1}}}

\newcommand{\eq}{eq.~}

\newcommand{\se}{sec.~}

\newcommand{\fig}{fig.~}

\newcommand{\tinymsbar}{{\overline{\mbox{\tiny\rm{MS}}}}}
\newcommand{\Lambdamsbar}{{\Lambda_\tinymsbar}}

\newcommand{\alphas}{\alpha_{\rm s}}

\newcommand{\Ns}{N_{\rm s}}
\newcommand{\Tc}{T_{\rm c}}
\newcommand{\rO}{r^{ }_0}
\newcommand{\tO}{t^{ }_0}
\newcommand{\betac}{\beta^{ }_c}
\newcommand{\Nt}{N^{ }_\tau}

\newcommand{\rmO}{{\mathcal{O}}}

\def\lsi{\raise0.3ex\hbox{$<$\kern-0.75em\raise-1.1ex\hbox{$\sim$}}}
\def\gsi{\raise0.3ex\hbox{$>$\kern-0.75em\raise-1.1ex\hbox{$\sim$}}}
\newcommand{\lsim}{\mathop{\lsi}}

\newcommand{\rmii}[1]{{\mbox{\tiny\rm{#1}}}}

\newcommand{\re}{\mathop{\mbox{Re}}}

\newcommand{\Tint}[1]{{\hbox{$\sum$}\!\!\!\!\!\!\!\int\,}_{\!\!\!\!\raise-0.9ex\hbox{$\scriptstyle{#1}$}}}
\newcommand{\Tinti}[1]{{{\Sigma}\!\!\!\!\raise0.3ex\hbox{$\int$}_\rmii{${#1}$}}}

\newcommand{\bi}{\begin{itemize}}
\newcommand{\ei}{\end{itemize}}
\newcommand{\hide}[1]{ }

\makeatletter
\renewcommand\section{\@startsection {section}{1}{\z@}%
                                   {-5.5ex \@plus -1ex \@minus -.2ex}% bfr-
                                   {2.3ex \@plus.2ex}%
                                   {\normalfont\large\bfseries}}
\renewcommand\subsection{\@startsection{subsection}{2}{\z@}%
                                     {-3.25ex\@plus -1ex \@minus -.2ex}%
                                     {1.5ex \@plus .2ex}%
                                     {\normalfont\normalsize\bfseries}}
\renewcommand\thesection {\@arabic\c@section}
\renewcommand\thesubsection   {\thesection.\@arabic\c@subsection}
\renewcommand{\@seccntformat}[1]{%
\csname the#1\endcsname.\hspace{1.0em}}
\makeatother
\begin{document}

\flushbottom

\begin{titlepage}

\begin{flushright}
BI-TP 2015/06 % Notes M.L. \\  % 1503.05652
\vspace*{1cm}
\end{flushright}
\begin{centering}
\vfill

{\Large{\bf
 Critical point and scale setting in SU(3) plasma: An update
}} 

\vspace{0.8cm}

A.~Francis$^{\rm a}$, 
O.~Kaczmarek$^{\rm b}$, 
M.~Laine$^{\rm c}$, 
T.~Neuhaus$^{\rm d}$ 
 and 
H.~Ohno$^{\rm e,f}$

\vspace{0.8cm}

$^\rmi{a}$%
{\em
        Department of Physics and Astronomy, 
        York University, 4700 Keele St., \\
        Toronto, ON M3J1P3, Canada\\
}

\vspace{0.3cm}

$^\rmi{b}$%
{\em
       Faculty of Physics, University of Bielefeld, 
        33501 Bielefeld, Germany\\
}

\vspace{0.3cm}

$^\rmi{c}$%
{\em
 Institute for Theoretical Physics, 
 Albert Einstein Center, University of Bern, \\ 
 Sidlerstrasse 5, 3012 Bern, Switzerland\\
}

\vspace*{0.3cm}

$^\rmi{d}$%
{\em
       Institute for Advanced Simulation,  
       J\"ulich Supercomputing Centre, \\
       52425 J\"ulich, Germany\\
}

\vspace*{0.3cm}

$^\rmi{e}$%
{\em
        Center for Computational Sciences, University of Tsukuba, 
        Tsukuba, \\ Ibaraki 305-8577, Japan\\
}

\vspace*{0.3cm}

$^\rmi{f}$%
{\em
        Physics Department, Brookhaven National Laboratory, 
        Upton, NY 11973, USA\\
}

%        \email{francis@kph.uni-mainz.de}, 
%        \email{okacz@physik.uni-bielefeld.de}, 
%        \email{laine@itp.unibe.ch},
%        \email{mmueller@physik.uni-bielefeld.de},
%        \email{t.neuhaus@fz-juelich.de},  
%        \email{hono@quark.phy.bnl.gov}

\vspace*{0.8cm}

\mbox{\bf Abstract}
 
\end{centering}

\vspace*{0.3cm}
 
\noindent
We explore a method developed in statistical physics which has been argued 
to have exponentially small finite-volume effects, in order to determine the 
critical temperature $\Tc$ of pure SU(3) gauge theory close to the continuum 
limit. The method allows us to estimate the critical coupling $\betac$ of the 
Wilson action for temporal extents up to $\Nt \sim 20$ with $\lsim 0.1\%$ 
uncertainties. Making use of the scale setting parameters $\rO$ and 
$\sqrt{\tO}$ in the same range of $\beta$-values, these results lead to 
the independent continuum extrapolations $\Tc \rO = 0.7457(45)$ and 
$\Tc \sqrt{\tO} = 0.2489(14)$, with the latter originating from a  
more convincing fit. Inserting a conversion  of $r_0$ from literature 
(unfortunately with much larger errors) yields 
$\Tc / \Lambdamsbar = 1.24(10)$. 

\vfill

%% %\noindent
%% %PACS numbers: 
%% %11.10.Wx, %        Finite temperature field theory
%% { %11.15.Ha, %        Lattice gauge theory } 
%% %12.38.Bx, %        Perturbative calculations in QCD
%% %12.38.Mh, %        Quark--gluon plasma
%% %14.40.Nd, %        Bottom mesons
%% %\\
%% %Keywords: Thermal Field Theory, Neutrino Physics, Resummation
 
\vspace*{1cm}
  
\noindent
April 2015

\vfill

\end{titlepage}

%%%%%%%%%%%%%%%%%%%%%%%%%%%%% SECTION %%%%%%%%%%%%%%%%%%%%%%%%%%%%%%%%%%%%
%
\section{Introduction}

Even though light quarks play an essential role for the phenomenological
understanding of heavy ion collision experiments it can be argued that,  
due to their large multiplicity in the initial state and their 
Bose-enhanced distribution functions in the plasma phase, gluons are 
the single most important degree of freedom influencing the formation
and evolution of QCD matter. 
Gluons are also much easier to study with non-perturbative
lattice methods than light quarks. 
Therefore studies of pure SU(3) gauge theory at high temperature 
continue to constitute an important laboratory system, 
both for developing numerical techniques 
and for gaining physics understanding on observables where a high
precision is needed. Recent examples of topics studied 
include scale setting, renormalization, and methods for 
statistical 
error reduction (cf.\ e.g.\ refs.~\cite{fixedscale}--\cite{precision2}). 
Our own interest stems from 
attempts to measure real-time observables such as transport 
coefficients~\cite{ding1}--\cite{thomas}, 
in which case theoretically well-founded methods~\cite{cuniberti} 
can probably be applied  (if at all) 
only after the infinite volume and  continuum limits have been 
reached with a high precision~\cite{cond}.

In the present contribution we use the pure SU(3) gauge theory as a test
bench for studying finite-volume scaling in the vicinity of 
a first-order phase transition.
Concretely, our primary goal is to determine the critical coupling $\betac$ 
for values of $\Nt$ much larger than have been achieved before 
(here $\Nt \equiv 1/(aT)$ is the number of lattice 
points in the periodic imaginary-time direction; $a$ is the lattice 
spacing; and $T$ is the temperature). Let us remark that values
of $\betac$ as a function of $\Nt$ have attracted recent interest 
as tests of semi-analytic models~\cite{pos12,pos13}, and indeed new 
high-precision values at large $\Nt$ put the functional dependences 
predicted by these frameworks under tension~\cite{first}. 

The second focus point of our study is that of 
scale setting~\cite{scale}. In particular, we consider two scales that
have been frequently employed, denoted by $\rO$~\cite{r0} 
and $\sqrt{\tO}$~\cite{t0}. Neither of these scales has a direct
physics interpretation; however they are relatively straightforward
to measure and can in principle
be related to physical quantities in a separate study. On the other
hand, in the thermal context there is one directly physical
quantity, the critical temperature $\Tc$, which would have 
certain advantages as a scale setting parameter, 
permitting for instance for an easy
comparison of theories with 
different matter contents but with similar
macroscopic properties (this assumes, of course, 
that all theories considered have a sharply defined transition point). 
Therefore, we make use of our results in order to obtain 
a largely independent estimate for $\Tc \rO$~\cite{sn} and
a new estimate for $\Tc \sqrt{\tO}$. It should be acknowledged, however, 
that close to the continuum limit we also see indications of growing
systematic uncertainties, particularly in case of $\rO$.

The plan of this note is the following. 
After introducing and testing the basic method
of our study in \se\ref{se:method}, we employ it in order
to estimate the critical coupling $\betac$ as a function
of $\Nt$ in \se\ref{se:betac}. 
The issue of scale setting is addressed 
in \se\ref{se:continuum}, and we conclude
in \se\ref{se:concl}.

%%%%%%%%%%%%%%%%%%%%%%%%%%%%% SECTION %%%%%%%%%%%%%%%%%%%%%%%%%%%%%%%%%%%%
%
\section{Method}
\la{se:method}

%%%%%%%%%%%%%%%%%%%%%%%%%%%%% SUBSECTION %%%%%%%%%%%%%%%%%%%%%%%%%%%%%%%%%
%
% \subsection{General idea}

The Wilson plaquette action, 
\be
 S^{ }_\rmii{W}
 \equiv
 \frac{\beta}{6}
 \sum_{x,\mu,\nu} \tr (\mathbbm{1} - P^{ }_{\mu\nu})
 \;, 
 \la{S}
\ee
studied on an $\Nt \times N_s^3$ lattice with periodic boundary conditions
in all directions, has a global Z(3) symmetry that is broken at and above 
the transition point for $\Ns\to\infty$. We denote the location
of the transition point by $\betac$. Theoretical
arguments~\cite{sy} and
empirical evidence~\cite{old1}
suggest that this is a first order phase transition. 

It has been shown through a study of $q$-state Potts models
in three dimensions~\cite{bkm,bj} that even though most 
observables, such as susceptibilities, show powerlike 
finite-volume effects at a first-order transition point, there is a particular
definition of a pseudocritical point for which 
finite-volume effects are exponentially suppressed. This is 
obtained if the ``weights'' of the phases with no degeneracy
($w_c$) and with $q$-fold degeneracy ($w_d$) are related through 
\be
 q\, w_c = w_d
 \;.
\ee
The weight can be defined through the ``volume'' of the distribution
of some observable which has a good overlap with the order parameter. 
More formally, the weight corresponds to the partition function
associated with the phase considered. 

%%%%%%%%%%%%%%%%%%%%%%%%%%%%% FIGURE %%%%%%%%%%%%%%%%%%%%%%%%%%%%%%%%%%%%%%%
%
\begin{figure}[t]
    \vspace*{+4.5cm}
    \hspace*{-4mm}% 
  \begin{minipage}[t]{5.0cm}
    \vspace*{-5cm}
    \hspace*{-3mm}\epsfig{file=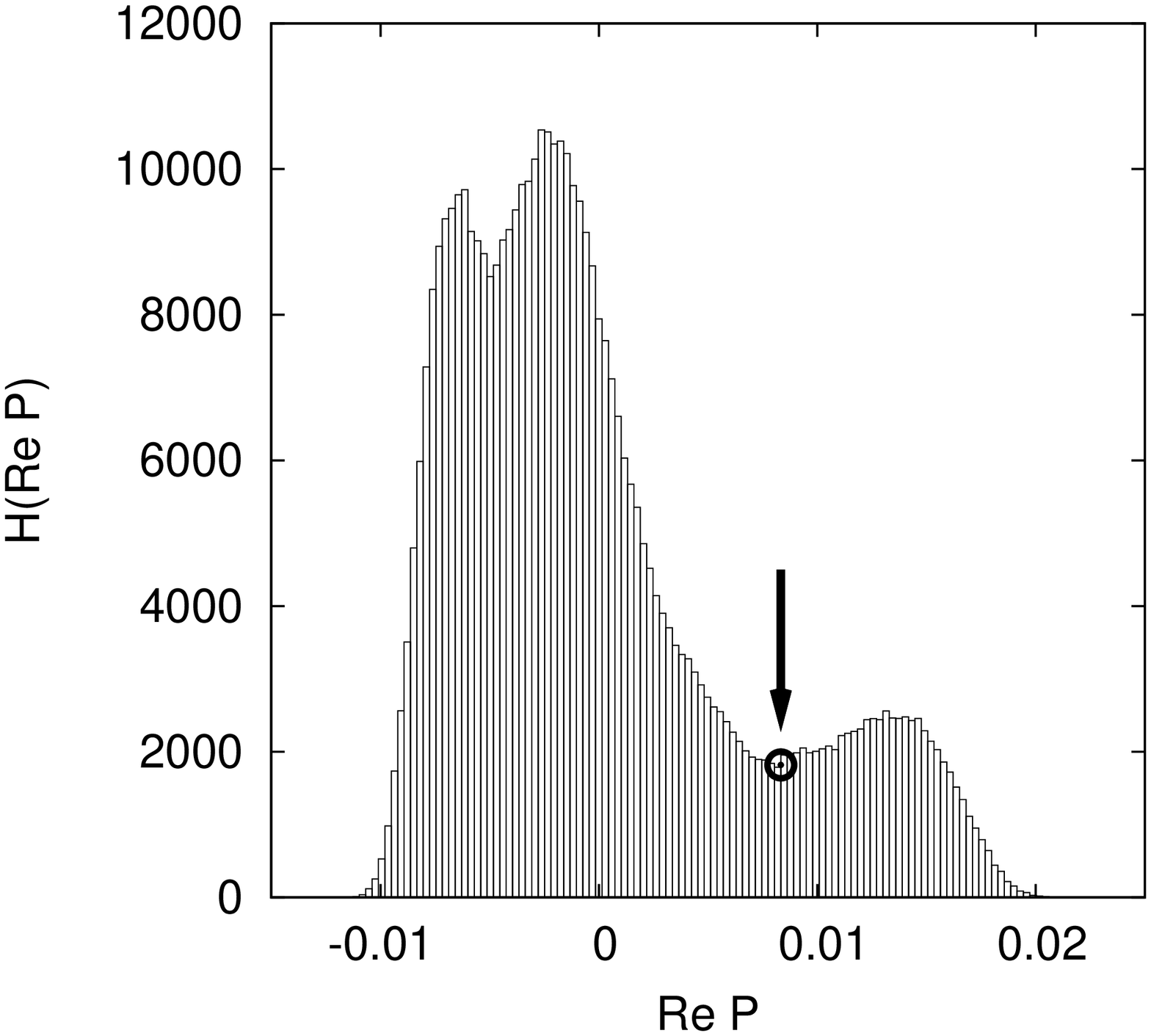,angle=270,width=6.3cm}
  \end{minipage}
  \begin{minipage}[t]{5.0cm}
    \vspace*{-5.4cm}
    \hspace*{-3mm}\epsfig{file=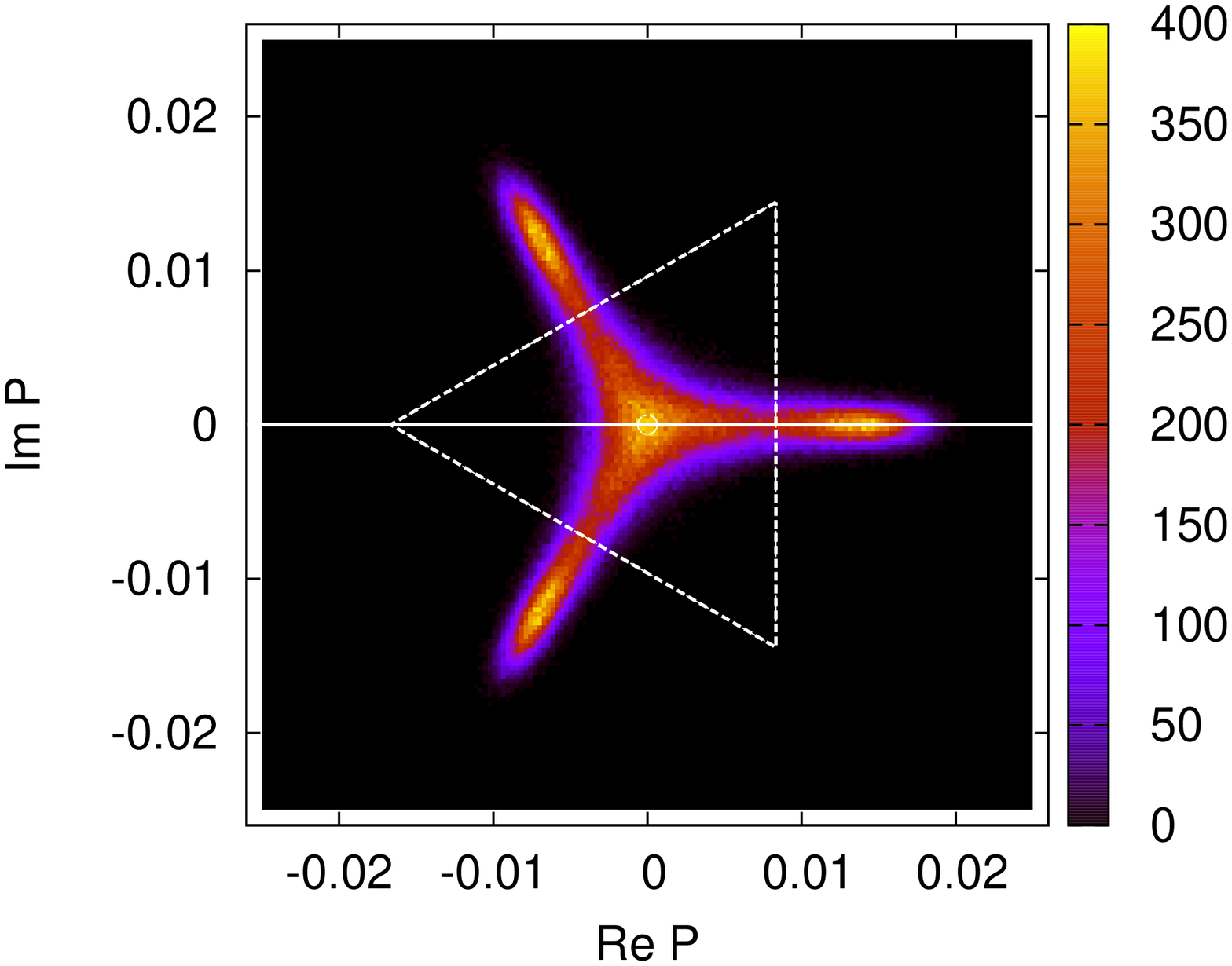,angle=270,width=6.9cm}
  \end{minipage}
  \begin{minipage}[t]{5.0cm}
    \vspace*{-5.4cm}
    \hspace*{2mm}\epsfig{file=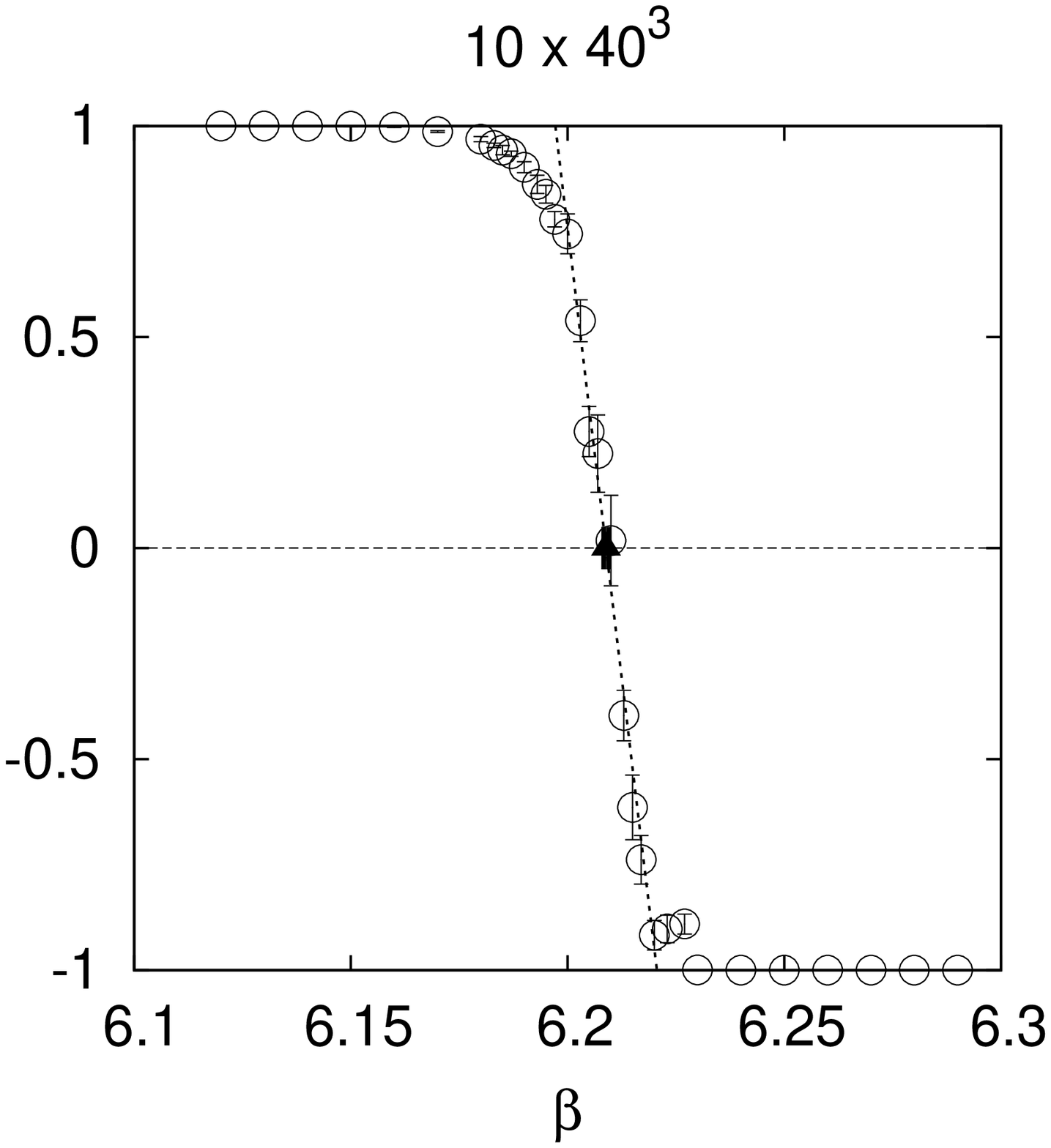,angle=270,width=5.1cm}
  \end{minipage}
    \vspace*{-0.5cm}
\caption[a]{
 Left: Determination of the right-most minimum (open circle) 
       from the distribution
       of $\re {P} $. 
 Middle: The corresponding triangle separating 
       the two phases, with the vertical line placed at the position
       of the open circle. 
 Right: The resulting function $s(\beta)$
        (cf.\ \eq\nr{theta}), permitting for an
        estimate of $\betac$ from the crossing of zero.
        The statistics of each data point is 
        $\rmO(10^5)$ sweeps; statistical errors 
        are based on jackknife estimates.   
 }
\label{fig:ill}
\end{figure}
%
%%%%%%%%%%%%%%%%%%%%%%%%%%%%%%%%%%%%%%%%%%%%%%%%%%%%%%%%%%%%%%%%%%%%%%%%%%%%%

For SU(3), a suitable observable is 
the Polyakov loop expectation value. Carrying 
out measurements in the vicinity of $\betac$, we define
\be
 s(\beta) \;\equiv\; \frac{3 w_c - w_d}{3 w_c + w_d}
 \;. \la{theta}
\ee
By construction $s(\beta)$ equals $+1$ deep in the confined
and $-1$ deep in the deconfined phase. 
The critical point is obtained by 
interpolating to the location where $s(\betac) = 0$.

%%%%%%%%%%%%%%%%%%%%%%%%%%%%% SUBSECTION %%%%%%%%%%%%%%%%%%%%%%%%%%%%%%%%%
%
% \subsection{Practical test}
% \la{ss:test}

%%%%%%%%%%%%%%%%%%%%%%%%%%%%%%%% FIGURE %%%%%%%%%%%%%%%%%%%%%%%%%%%%%%%%%%%%
%
\begin{figure}[t]

\begin{center}
  \hspace*{-3mm}\epsfig{file=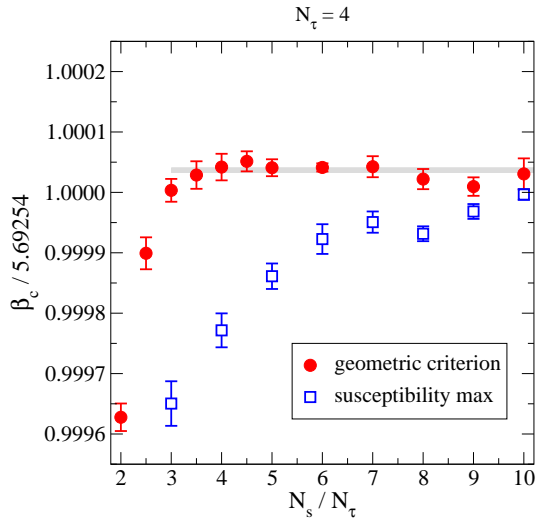,width=7.0cm}%
%  ~~~\epsfig{file=Nt6.eps,width=7.0cm}
\end{center}

\vspace*{-3mm}

\caption[a]{
       The pseudocritical couplings extracted from our method
       at $\Nt = 4$ (closed circles), 
       normalized to the central value of the infinite-volume estimate 
       $\betac = 5.69254(24)$ from ref.~\cite{old1}. 
       We also compare with susceptibility maxima from  
       ref.~\cite{bw} (open squares). The grey band illustrates our
       infinite-volume extrapolation (constant fit to $\Ns/\Nt > 3$). 
% Right: The same for $\Nt = 6$, 
%       normalized to the central value of 
%       $\betac = 5.8941(5)$ from ref.~\cite{old1}.
 }
\label{fig:demo}
\end{figure}
%
%%%%%%%%%%%%%%%%%%%%%%%%%%%%%%%%%%%%%%%%%%%%%%%%%%%%%%%%%%%%%%%%%%%%%%%%%%%%%

In order to implement the idea, we need to introduce a criterion
for separating a distribution into contributions from different phases. 
In a finite volume, when the distributions overlap, the procedure is 
not unique. In this study, 
we define a separatrix by looking for a minimum
in the distribution of $\re {P} $, where ${P}$
denotes the Polyakov 
loop (cf.\ \fig\ref{fig:ill}(left)). 
This minimum is employed for 
defining a triangle separating the two phases 
(cf.\ \fig\ref{fig:ill}(middle)). 
The resulting weights are the inputs for \eq\nr{theta}; $\betac$
is obtained by a linear interpolation from points on both sides
of the zero (cf.\ \fig\ref{fig:ill}(right)). 

The results obtained with this procedure are shown
in \fig\ref{fig:demo} for 
$\Nt = 4$. 
They have been normalized
to a classic value from ref.~\cite{old1}, and are compared with 
recent high-precision pseudocritical points extracted from  
Polyakov loop susceptibility maxima~\cite{bw}. 
We conclude that for $\Ns > 3 \Nt$,   
no finite-volume effects can be observed within our resolution
($\sim 0.005$\%). For $\Ns < 3 \Nt$, we expect $\betac$ to be slightly
underestimated.

%%%%%%%%%%%%%%%%%%%%%%%%%%%%% SECTION %%%%%%%%%%%%%%%%%%%%%%%%%%%%%%%%%%%%
%
\section{Results at finite lattice spacing}
\la{se:betac}

We carried out measurements
for $4 \le \Nt \le 22$, increasing $\Nt$ in steps of two. 
We computed on several volumes for ensembles with $\Nt \le 18$, 
verifying that volume dependence is below statistical 
uncertainties. Subsequently we fit the data at $\Ns > 3 \Nt$ to a constant. 
Given the resources at our disposal we used
a single spatial extent $\Ns = 64$ for $\Nt = 20,22$. Here, minor 
finite-volume effects start to contaminate our results. 
However, based on \fig\ref{fig:demo}, we expect the 
effects from a simulation
with $\Ns/\Nt = 64/22 = 2.9$ to be below the 0.01\% level, 
thereby being much below statistical errors. In contrast,  
at the smallest $\Nt$
where statistical errors are extremely small, we have  
artificially saturated the errors at a constant value $\sim 0.005\%$, 
corresponding to the expected uncertainty from finite-volume effects. 
Our final results at fixed $\Nt$, 
together with previous estimates from the literature, 
are collected in table~\ref{table:betac}.

%%%%%%%%%%%%%%%%%%%%% TABLE %%%%%%%%%%%%%%%%%%%%%%%%%%%%%%%%%%%%%
%
\begin{table}[t]

\small{
\begin{center}
\begin{tabular}{llll|lll}
 \hline
 $\Nt$ & $\betac$~\cite{old1,old2} 
  & $\betac$~\cite{bw} 
  & $\betac$~\cite{old3}
  & $\betac$~[our value]
  & $\Ns$ used & $N_\rmi{total}$ \\[1mm]
 \hline 
 4 & 5.69254(24) & 5.692469(42) &           & 5.69275(28) & 14,...,40 &
   83$\times 10^6$ \\
 5 &             &              & 5.8000(5) &   &  & \\
 6 & 5.8941(5)   & 5.89410(11)  &           & 5.89425(29) & 20,...,40 &
   28$\times 10^6$ \\
 8 & 6.0624(10)  & 6.06212(44)  &           & 6.06239(38) & 28,32 &
   4.2$\times 10^6$ \\
 10 &            &              &           & 6.20873(47) & 32,...,56 &
   15$\times 10^6$ \\
 12 & 6.3380(17) &              &           & 6.33514(45) & 40,...,72 &
   21$\times 10^6$ \\
 14 &            &              &           & 6.4473(18) & 48,56 &
   12$\times 10^6$ \\
 16 &            &              &           & 6.5457(40) & 64 &
   2.5$\times 10^6$ \\
 18 &            &              &           & 6.6331(20) & 56,64 &
   3.6$\times 10^6$ \\
 20 &            &              &           & 6.7132(26) & 64 &
   4.0$\times 10^6$ \\
 22 &            &              &           & 6.7986(65) & 64 &
   5.9$\times 10^6$ \\
 \hline 
\end{tabular} 
\end{center}
}

\caption[a]{
  The infinite-volume critical points of SU(3) gauge theory
  according to various studies. $N_\rmi{total}$ indicates the 
  total numbers of configurations 
  (all volumes and values of $\beta$).
  Our data are based on constant fits to $\Ns > 3 \Nt$
  whenever several volumes are available. For 
  $\Nt = 4,6$ we have artificially enlarged the errors to account
  for systematics related to exponentially small volume
  corrections (cf.\ the text). 
 }
\label{table:betac}
\end{table}
%
%%%%%%%%%%%%%%%%%%%%%%%%%%%%%%%%%%%%%%%%%%%%%%%%%%%%%%%%%%%%%%%%%%%%%

%%%%%%%%%%%%%%%%%%%%%%%%%%%%% SECTION %%%%%%%%%%%%%%%%%%%%%%%%%%%%%%%%%%%%
%
\section{Continuum extrapolations}
\la{se:continuum}

In this section we convert the lattice-specific numbers of 
table~\ref{table:betac} to values of $\Tc$ in 
physical units. In order to achieve this two different scale setting
parameters are considered, $\rO$ and $\sqrt{\tO}$, with the
latter leading to a noticeably better description of the thermal 
data (cf.\ \se\ref{ss:t0}).

%%%%%%%%%%%%%%%%%%%%%%%%%%%%% SUBSECTION %%%%%%%%%%%%%%%%%%%%%%%%%%%%%%%%%
%
\subsection{Scale $\rO$}
\la{ss:r0}

%%%%%%%%%%%%%%%%%%%%% TABLE %%%%%%%%%%%%%%%%%%%%%%%%%%%%%%%%%%%%%
%
\begin{table}[t]

\small{
\begin{center}
\begin{tabular}{lll|llcc}
    	\hline
        $\beta$ & $\rO/a$~\cite{pos18} & $\rO/a$~\cite{ns} & 
        $\rO/a$~[our value] &  
    	$\Nt\times \Ns^3$ & $N_\rmi{conf}$ \\[1mm] 
    	\hline
   5.7  &    2.922(9) & \\
   5.8  &    3.673(5) & \\
   5.95 &    4.898(12) & \\
   6.07 &    6.033(17) & \\
   6.2 &    7.380(26) & \\
  6.3 & & & 8.52(4) & $32 \times 32^3$ & 216   \\ 
  6.3 & & & 8.51(2) & $32 \times 48^3$ & 211  \\ 
  6.3 & & & 8.52(2)$^\star$ & $32 \times 64^3$ & 202  \\ 
  6.336 & & & 8.95(3) & $64 \times 32^3$ & 220  \\ 
  6.4 & & & 9.80(3) &  $36 \times 36^3$ & 206  \\ 
  6.5 & & & 11.16(2) &  $44 \times 44^3$ & 202  \\ 
  6.57 & & 12.18(10)$^{\star\star}$ \\
  6.69 & & 14.20(12)$^{\star\star}$ \\
  6.81 & & 16.54(12)$^{\star\star}$ \\
  6.92 & & 19.13(15)$^{\star\star}$ \\
    	\hline
\end{tabular}
\end{center}
}

\caption[a]{
  The results for $\rO/a$ that have been used in our analysis. 
  For $\beta = 6.3$ only the largest volume (indicated with an
  asterisk) has been included in subsequent fits. 
  The values from ref.~\cite{ns}, marked with a double asterisk, 
  do not come directly from $\rO$ but rather another scale $r_c$, 
  which has been converted into $\rO$ through a continuum relation, 
  whose systematic uncertainties are included 
  in the errors. 
 }
\label{table:r0}
\end{table}
%
%%%%%%%%%%%%%%%%%%%%%%%%%%%%%%%%%%%%%%%%%%%%%%%%%%%%%%%%%%%%%%%%%%%%%

The scale $r_0 / a$~\cite{r0} has been measured as a function of $\beta$
in refs.~\cite{pos18,ns}
(see ref.~\cite{uh} and references therein for previous work). 
We complement these results by
a new set of simulations, with parameter values and results listed in 
table~\ref{table:r0}. The measurements
were separated by 500 heatbath-overrelaxation updates.
A number of standard techniques 
for statistical error reduction~\cite{st1,st2,st3} 
were implemented in order to obtain these results. 
The static potential is extracted from Wilson loops with 
an ansatz based on two exponentials. The distance appearing
in the static potential is tree-level improved~\cite{ns},
and subsequently a B-spline interpolation is carried out
in order to extract $\rO/a$ from its definition~\cite{r0}. 
(Note that due to the several steps involved, measurements are costly 
and systematic errors are difficult to get fully under control,
particularly at large $\beta$.) 

%%%%%%%%%%%%%%%%%%%%% TABLE %%%%%%%%%%%%%%%%%%%%%%%%%%%%%%%%%%%%%
%
\begin{table}[t]

\small{
\begin{center}
\begin{tabular}{llllll}
    	\hline
    	$\beta$ & $(t_0/a^2)^{\rm Wilson}$ 
        & $(t_0/a^2)^{\rm Wilson~imp.}$ & $(t_0/a^2)^{\rm Clover}$
        & $\Nt\times \Ns^3$ & $N_\rmi{conf}$ \\[1mm]
    	\hline
    5.6923 & 0.6109(10) & 0.8234(9) & 1.0124(11) & $16 \times 16^3$ & 455 \\
    5.6923 & 0.6103(7)  & 0.8229(6) & 1.0119(7) & $16 \times 24^3$ & 313 \\
    5.6923 & 0.6095(5)  & 0.8220(5) & 1.0104(6) & $16 \times 32^3$ & 248 \\
    5.6923 & 0.6010(4)  & 0.8226(4) & 0.9905(4) & $24 \times 32^3$ & 233 \\
    5.6923$^\star$ 
           & 0.6097(3)  & 0.8223(3) & 0.9800(4) & $32 \times 32^3$ & 221 \\
   \hline
    5.8941 & 1.9520(22) & 2.0989(22) & 2.2889(24) & $24 \times 24^3$ & 465 \\
    6.0625 & 3.7129(39) & 3.8507(39) & 4.0626(41) & $32 \times 32^3$ & 673 \\
    6.2083 & 5.9521(65) & 6.0873(66) & 6.3284(68) & $40 \times 40^3$ & 476 \\
    6.3352 & 8.668(11)  & 8.802(11)  & 9.076(12)  & $48 \times 48^3$ & 315 \\
    6.4487 & 11.958(18) & 12.091(18) & 12.397(18) & $56 \times 56^3$ & 254 \\
    6.5509 & 15.769(23) & 15.901(23) & 16.240(24) & $64 \times 64^3$ & 305 \\ 
    6.7130 & 24.222(35) & 24.355(35) & 24.752(36) & $80 \times 80^3$ & 250 \\
    	\hline
\end{tabular}
\end{center}
}

\caption[a]{
  Our results for $\tO/a^2$. The $\beta$-values
  correspond approximately to those in table~\ref{table:betac}
  (apart from $\Nt=18,22$), with 
  $\Nt$ scaled up by a factor 4 in each case. For $\beta = 5.6923$ 
  only the largest volume (indicated with an asterisk) has been included
  in subsequent fits. 
 % At $\beta \gsim 6.4$ systematic errors could
 % be larger than statistical ones (cf.\ the text). 
 }
\label{table:t0}
\end{table}
%
%%%%%%%%%%%%%%%%%%%%%%%%%%%%%%%%%%%%%%%%%%%%%%%%%%%%%%%%%%%%%%%%%%%%%

In order to permit for a subsequent interpolation, our data and older
values~\cite{pos18,ns} are fit
in the range $\beta\in(5.7,6.92)$ to a rational ansatz
inspired by ref.~\cite{pos20}: 
\be
 \ln\Bigl( \frac{\rO}{a} \Bigr)
 = 
 \biggl[\frac{\beta}{12 b_0} + \frac{b_1}{2b_0^2}
  \ln\Bigl( \frac{6 b_0}{\beta} \Bigr)  \biggr]
 \frac{1 + c_1 /\beta + c_2 / \beta^2}
      {1 + c_3/\beta + c_4 / \beta^2}
 \;, \la{duerr}
\ee
where $b_0 \equiv 11/(4\pi)^2$ and $b_1 \equiv 102/(4\pi)^4$.
The fit parameters obtained read\footnote{%
 For the sake of reproducibility of subsequent results we show
 more digits than are statistically significant. 
 } 
\be
 c_1 = -8.17273 \;, \quad
 c_2 = 14.9600  \;, \quad
 c_3 = -3.95983 \;, \quad
 c_4 = -5.30334 \;, \quad
 \chi^2/{\rm d.o.f.} = 0.7
% c_1 = -8.16094 \;, \quad
% c_2 = 14.8968  \;, \quad
% c_3 = -3.94079 \;, \quad
% c_4 = -5.39027 \;, \quad
% \chi^2/{\rm d.o.f.} = 0.8
 \;. \la{interp}
\ee
Based on the above equation, we convert the results in 
table~\ref{table:betac} to values of $\rO \Tc$:
 $\rO \Tc = (\rO / a)(\betac) /  \Nt$. 
Subsequently we perform the 
extrapolation $(a/\rO)^2 \to 0$ using a fit quadratic 
in $(a/\rO)^2$, 
illustrated in \fig\ref{fig:r0Tc}(left), with the result
\be
 \rO \Tc = 0.7457(45) \;, \quad    %%% (6)
  \chi^2/{\rm d.o.f.} = 6.7
 \;. \la{Tcr0}
\ee
The error includes a rough estimate of systematic effects, 
encompassing the central values obtained by 
replacing the representation in \eq\nr{duerr} through
 $\ln(\rO/a) = \sum_{n=0}^3 a_n (\beta-6.25)^n$; by 
carrying out the continuum extrapolation with a cubic 
fit; and by omitting $\betac$ corresponding to $\Nt = 4$. 
The first method increases the central value 
($\Tc {\rO} \simeq 0.7496$), the second and third decrease it
($\Tc {\rO} \simeq 0.7412, 0.7424$, respectively). 
However, in the first case 
the quality of the continuum fit decreases further 
from the already poor one in \eq\nr{Tcr0}, 
whereas in the second case 
the scatter of the data in \fig\ref{fig:r0Tc}(left)
suggests that including too much
freedom in the fit distorts the outcome. 
A possible reason for the poor description
of the data close to the continuum limit 
could be that estimates of $\rO/a$ at $\beta > 6.4$ are 
systematically on the low side (by $\sim \rmO(1$\%)), but unfortunately 
we have not been able to confirm this suspicion.

The result in \eq\nr{Tcr0} can be compared with  
$\rO \Tc \simeq 0.7470(7)$
obtained in ref.~\cite{first}
based on peak positions of Polyakov loop susceptibilities
(here only statistical errors were included)\footnote{%
 For fixed $\Nt$ the results of ref.~\cite{first} are consistent
 with the present ones, however their uncertainties from finite-volume
 effects are larger and only values up to $\Nt = 16$ could be reached. 
 Therefore systematic errors would be larger than in the present study
 (but are more difficult to estimate reliably). 
 }, 
as well as with the earlier value $\rO \Tc = 0.7498(50)$~\cite{sn}. 

Finally, we recall that e.g.\ the values 
$\rO \Lambdamsbar = 0.586(48)$~\cite{ns}, 
$\rO \Lambdamsbar = 0.602(48)$~\cite{sc}, 
$\rO \Lambdamsbar = 0.614(6)$~\cite{tad}, and
$\rO \Lambdamsbar = 0.637(32)$~\cite{pos23}
can be found in the literature (the third relies on the applicability
of tadpole-improved lattice perturbation theory and the fourth of 
continuum perturbation theory at hadronic scales). 
Using the second value yields
$\Tc/\Lambdamsbar = 1.24(10)$. Unfortunately the error is 
dominated by that in the relation of $\rO$ and $\Lambdamsbar$, 
so our new result in \eq\nr{Tcr0}
does not help to improve on previous estimates.  

%%%%%%%%%%%%%%%%%%%%%%%%%%%%% SUBSECTION %%%%%%%%%%%%%%%%%%%%%%%%%%%%%%%%%
%
\subsection{Scale $\sqrt{\tO}$}
\la{ss:t0}

The scale $\sqrt{\tO}$ is defined through the time that it takes for 
Wilson flow to adjust a chosen observable ($\equiv E$) to a 
predefined value~\cite{t0}. We measured ${\tO}$ for a number 
of $\beta \simeq \betac$, as listed in table~\ref{table:betac}. 
To study possible systematic effects, we made use of three 
different implementations of $E$, based on the standard plaquette, 
tree-level improved, and clover discretizations, all of which are
available within the DD-HMC package~\cite{ddhmc}. Like for $\rO$, 
the measurements
were separated by 500 heatbath-overrelaxation updates; the volumes
and the numbers of configurations used for measurements 
are shown in table~\ref{table:t0}.

%%%%%%%%%%%%%%%%%%%%%%%%%%%%%%%% FIGURE %%%%%%%%%%%%%%%%%%%%%%%%%%%%%%%%%%%%
%
\begin{figure}[t]

\begin{center}
  \epsfig{file=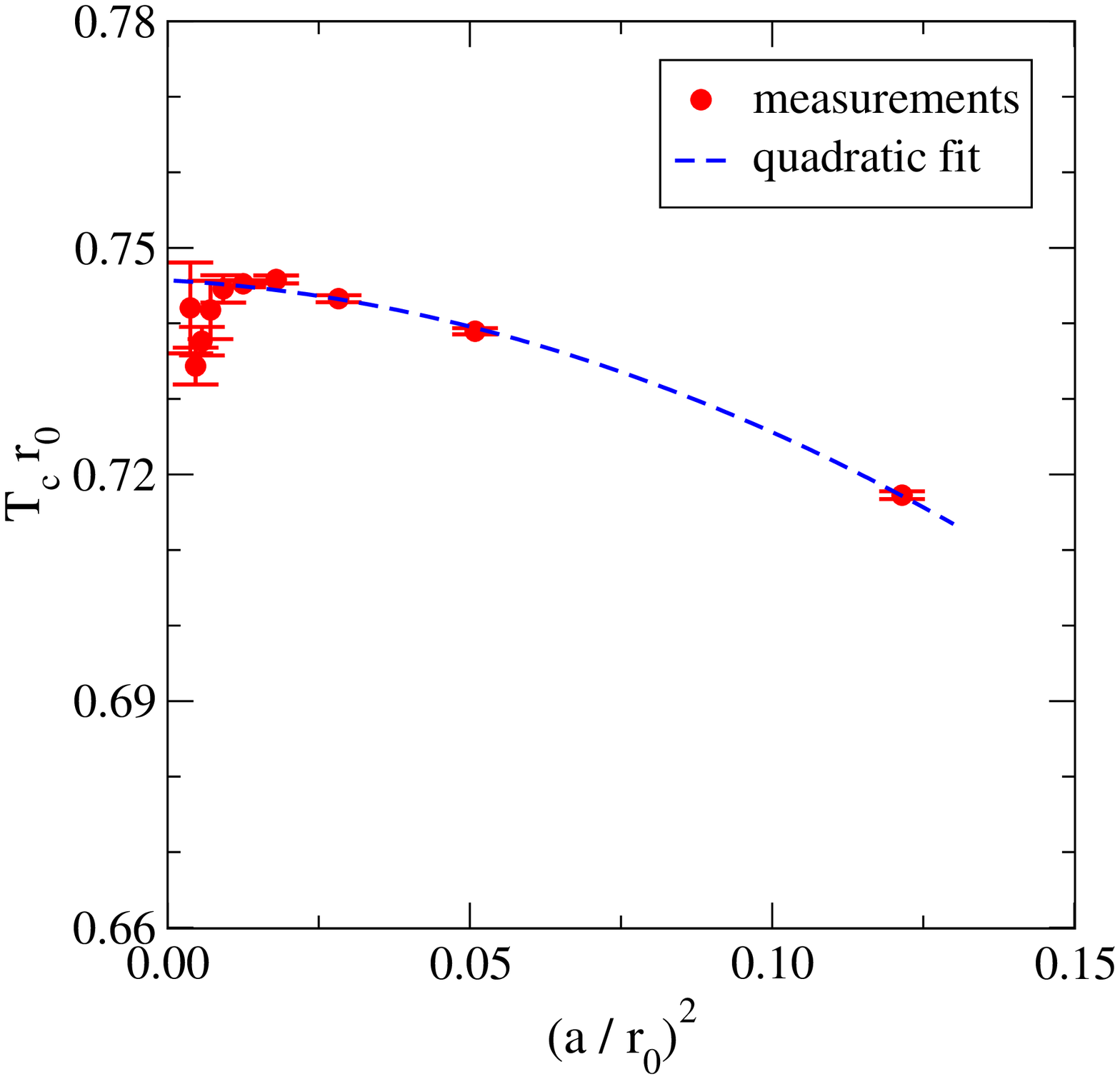,width=7.0cm}%
  \epsfig{file=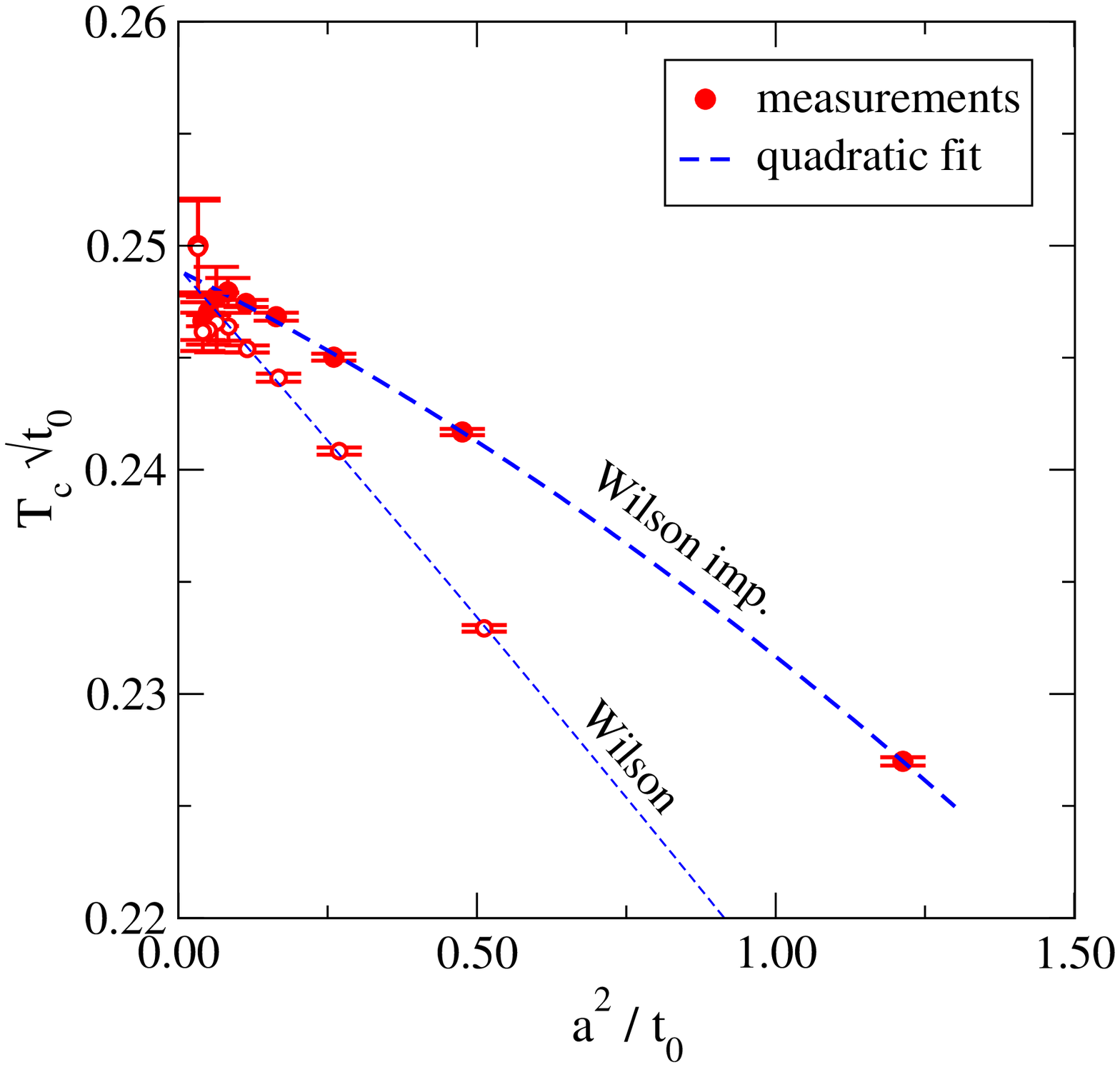,width=7.1cm}
\end{center}

\vspace*{-3mm}

\caption[a]{
 Left: Continuum extrapolation of $\Tc\rO$, 
 based on the data in table~\ref{table:betac} and the interpolation from 
 \eq\nr{interp}.
 Right: Analogous results 
 for $\Tc \sqrt{\tO}$ from the ``Wilson'' (open circles) and 
 ``Wilson imp.'' (closed circles) discretizations, 
 interpolated according to \eq\nr{interp2}.}
 \label{fig:r0Tc}
\end{figure}
%
%%%%%%%%%%%%%%%%%%%%%%%%%%%%%%%%%%%%%%%%%%%%%%%%%%%%%%%%%%%%%%%%%%%%%%%%%%%

Given that the $\beta$ values of table~\ref{table:betac} 
correspond to the critical point, a set of fixed 
physical volumes can be chosen
by scaling the corresponding $\Nt$ by a constant
amount. Setting $\Ns = 4 \Nt$ we ensure
that the box size is $L = 4/\Tc \simeq 5.3 \rO$. 
For the smallest $\beta$ we have carried
out test simulations also at larger volumes, finding consistent
results apart from the ``clover'' discretization for which 
volume dependence on the 3\% level 
is visible. For our final results we quote only those obtained with the
two variants of the 
``Wilson'' discretization that did not exhibit 
any volume dependence within statistical precision. 
Nevertheless systematic errors do grow with $\beta$, because a longer
integration trajectory in $t$ 
is needed and because autocorrelation times tend to grow. 

As before, we represent the data 
as in \eq\nr{duerr} for the interpolation, 
only this time replacing $\rO \to \sqrt{\tO}$.
The resulting parameters are 
(for the ``Wilson imp.'' discretization)\footnote{%
 For the sake of reproducibility of subsequent results we show
 more digits than are statistically significant. 
 } 
\be
 c_1 = -10.2116 \;, \quad
 c_2 = 25.6819 \;, \quad
 c_3 = -5.64462 \;, \quad
 c_4 = 2.26845 \;, \quad
 \chi^2/{\rm d.o.f.} = 2.3
 \;. \la{interp2}
\ee
With this interpolation the critical values in table~\ref{table:betac}
can be converted into $\Tc \sqrt{\tO}$; results are shown 
in \fig\ref{fig:r0Tc}(right). A fit quadratic in $a^2/\tO$ yields
\be
 \Tc \sqrt{\tO} = 0.2489(14)  \;, \quad  %%%% (4)
 \chi^2/{\rm d.o.f.} = 1.5
 \;. \la{Tct0}
\ee
The error bar here includes a rough estimate of systematic effects, 
encompassing  
the central values obtained by: 
(i) replacing ``Wilson imp.'' by ``Wilson'' or even the formerly
    excluded ``clover'' data; 
(ii) replacing the representation in \eq\nr{duerr} through
 $\ln(\sqrt{\tO}/a) = \sum_{n=0}^3 a_n (\beta-6.25)^n$; 
(iii) carrying out the continuum extrapolation with a cubic fit; 
(iv) omitting $\betac$ corresponding to $\Nt=4$ from the fit. 
The biggest deviations 
($\Tc \sqrt{\tO} \simeq 0.250$) result either from using ``clover'' data
which we assume to suffer from finite-volume effects, or from method (ii)
which leads to $\chi^2$ larger by more than 
an order of magnitude in \eq\nr{Tct0}. (An analysis
based on data for $\tO/a^2$ from previous literature can be found in 
ref.~\cite{thomas}, is however subject to noticeably larger finite-volume
effects than our current determination.)

Comparing \eq\nr{Tct0} with \eq\nr{Tcr0}, we extract
$\sqrt{\tO}/\rO  = 0.3338(28)$, in perfect agreement with 
$\sqrt{\tO}/\rO  = 0.3343(21)$ from refs.~\cite{t0,r0t0}.
It is comforting to find a good agreement from a largely 
independent analysis. 

%%%%%%%%%%%%%%%%%%%%%%%%%%%%% SECTION %%%%%%%%%%%%%%%%%%%%%%%%%%%%%%%%%%%%
%
\section{Conclusions}
\la{se:concl}

In this note
we have demonstrated that with modern resources and an opportune choice
of an observable, the critical coupling $\betac$ of the Wilson plaquette
action can be determined with $\lsim 0.1$\% errors up to $\Nt \sim 20$
(cf.\ table~\ref{table:betac}). Subsequently, 
the critical temperature $\Tc$ of pure SU(3) gauge
theory could serve as a valid scale setting parameter for values 
of the Wilson coupling in the range
$5.7 \lsim \beta \lsim 6.8$
(cf.\ table~\ref{table:betac}, from which the lattice spacing $a$
is obtained as $a = 1/(\Nt\Tc)$ if we simulate at the $\betac$
corresponding to $\Nt$). 
Unfortunately these values are not large enough 
for scale setting on the very fine lattices 
(for instance $\Nt = 48$, $\beta \simeq 7.8$) 
that are being used for studying transport 
observables close to the continuum limit~\cite{ding1}--\cite{thomas}. 
Therefore ``theoretical'' quantities like $\rO$ and $\sqrt{\tO}$
continue to be needed as intermediate steps. On this point 
our study suggests that, with comparable numerical effort, 
employing $\sqrt{\tO}$ may yield more stable results
than $\rO$, however being assured that
systematic errors are below the percent level remains a challenge 
for $\beta > 6.4$. If $\sqrt{\tO}$
is indeed used for scale setting, a conversion
to $\Tc$ can be carried out through \eq\nr{Tct0}:
$ \sqrt{\tO} \Tc = 0.2489(14)$. 

For various comparisons
of lattice data with continuum 
perturbation theory, it would be very welcome to improve
on our knowledge of $\sqrt{\tO}\Lambdamsbar$, whose uncertainty is 
currently an order of magnitude larger 
than that of $\sqrt{\tO}\Tc$.\footnote{%
 After the appearance of the eprint version of our paper 
 a study appeared in which a possible strategy for this task 
 was suggested~\cite{flow2}.
 }
Another issue worth further consideration is whether 
the method of \se\ref{se:method}, which relied on the 
breaking of a discrete symmetry, could be generalized
to the case of a continuous symmetry (such as a chiral symmetry).

%%%%%%%%%%%%%%%%%%%%%%%%%%%%% SECTION %%%%%%%%%%%%%%%%%%%%%%%%%%%%%%%%%%%%
%
\section*{Acknowledgments}                                    
                       
We thank M.~M\"uller for collaboration at initial stages of this project. 
Our work has been supported
in part by the DFG under grant GRK881, 
by the SNF under grant 200020-155935,
and by the European Union through HadronPhysics3
(grant 283286) and ITN STRONGnet
(grant 238353).
Simulations were
performed using JARA-HPC resources at the RWTH Aachen
(projects JARA0039 and JARA0108), JUDGE/JUROPA at the JSC J\"ulich, 
the OCuLUS Cluster at the Paderborn Center for Parallel
Computing, and the Bielefeld GPU cluster.

%%%%%%%%%%%%%%%%%%%%%%%%%%%%%%%%%%%%%%%%%%%%%%%%%%%%%%%%%%%%%%%%%%%%%%%%%%%
%

\end{document}